# Understanding the indirect effects of interactive systems within systems of systems

Laetitia Bornes laetitia.bornes@isae-supaero.fr
Catherine Letondal catherine.letondal@enac.fr
Rob Vingerhoeds rob.vingerhoeds@isae-supaero.fr

*Abstract*
Until recently, research into the sustainable design of interactive systems has primarily focused on the direct material impact of a system, through improving its energy efficiency and optimizing its lifecycle. Yet the way a system is designed and marketed often has wider repercussions, such as rebound effects, and systemic change in practices. These effects are harder to assess (and to anticipate) than the direct physical impact of the construction and use of the system itself. Current tools are unable to account for the complexity of these effects: the underlying causal mechanisms, their multi-level nature, their different temporalities, and the variety of their consequences (environmental and societal). This is why we are seeking to develop a specific methodology and tool, inspired by systemic design and system dynamics. These are intended for decision-makers and designers of interactive systems within systems of systems (for example, in the fields of agricultural robotics or public transportation). In this paper, we present this modeling approach and our prototype tool through the example of a second-hand clothing sales platform.

*Keywords*
Sustainability, Systems of Systems, Sociotechnical systems, Rebound effect, Systemic, Methodology, Modeling tool

*The indirect effects of an interactive system within a sociotechnical system*

In interactive systems[1] engineering, until recently, efforts to move towards a more sustainable future have mainly focused on the direct material impact of a system, by improving energy efficiency during its use and optimizing its lifecycle to reduce waste, pollution, and greenhouse gas emissions. However, the impact of an interactive system is not limited to the direct physical impact of its manufacturing, use, maintenance, and end of life (first order effect). In fact, the introduction of a new product (or a new technology) into society very often has indirect consequences, due to how it is used and the changes in societal practices it induces (Coroamă 2019). A simple example is the rebound effect, first identified by William Stanley Jevons in 1865 in relation to the steam train (Jevons 1865). The rebound effect occurs when the optimization of a system leads to a saving (in time or cost) which has the effect of increasing overall consumption.

In addition to direct (first order) effects, there are several indirect effects, including:
- **direct rebound (second order) effect**: A rebound effect where increased efficiency, associated cost reduction and/or convenience of a product or service results in its increased use because it is cheaper or otherwise more convenient. (ITU 2022)

Example: In the case of the car, improved engine efficiency enables drivers to save fuel: they can drive more often or for longer, and, for example, live further away from their work, which results in an overall increase in fuel consumption.
- **indirect rebound (second order) effect**: A type of rebound effect where savings from efficiency cost reductions enable more income to be spent on other products and services. (ITU 2022)

Example: Some drivers will spend their fuel savings on other activities, such as flying on holiday.
- **systemic (higher order) effect**: The indirect effect (including but not limited to rebound effects) other than first and second order effects occurring through changes in consumption patterns, lifestyles, and value systems. (ITU 2022)

---

[1] In this context, "interactive system" refers to a computerized system whose behavior adapts to users' actions.

Example: The introduction of the car has completely changed the way cities are organized and how people get around, to the extent that it is difficult to do without a car in certain regions or for certain professions.

Second order and higher order effects do not necessarily exceed efficiency gains and are not even necessarily negative. Nevertheless, it is imperative to understand them, as they are now perceived as potentially very impactful, in terms of intensity and duration, hence the recent interest in sustainability research (Coroamă 2020). These effects are on a much larger scale (the higher the order, the larger the scale), and are interwoven with social dynamics, practices, and lifestyles, making them very difficult to assess, let alone anticipate.

"Systemic effects have wider boundaries of analysis and are more difficult to quantify and investigate but are nonetheless very relevant" as stated by the IPCC (Pathak 2021) in the Working Group III report and described by Gauthier Roussilhe (2022).

*A 'quali-quantitative' modeling methodology*

Designers, decision-makers and policymakers lack the tools and methods to understand these effects and to visualize the dynamics of sociotechnical systems (systems of systems). The ITU-T L. I480 methodology proposes mapping these effects using a consequence tree (ITU 2022), which consists of listing the various possible effects and ranking them in the form of a tree. This representation does not shed light on certain feedback dynamics, when consequences can themselves become causes. In addition, this tool is qualitative and does not allow orders of magnitude to be represented. Yet many of the problems we face are physical and quantifiable ($CO_2$ emissions, depletion of resources, land artificialization, etc.). Other existing tools mainly focus on environmental impacts (particularly GHG) and ignore societal impacts (working conditions, access to healthcare, etc.). However, it is necessary to adopt a systemic approach if one intends to respect the planetary boundaries, as well as the social foundations, as described in the Doughnut model (Raworth 2017).

Drawing on system dynamics (Sterman 2000) and systemic design (Jones 2020), we are seeking to develop a 'quali-quantitative' modeling methodology (Bornes et al. 2022) and tool (Bornes 2023). To develop this methodology, we rely on the activities of Group Model Building (Bérard 2010), which we integrate into the interactive systems design process. We conducted interviews with systemic designers and continuously assess and refine this methodology by applying it to case studies. Specifically, we held two workshops with professional user experience designers on the case of a second-hand clothing platform (partially presented in the next section), and we also plan two new workshops with systemic designers (Bornes et al. 2023). Additionally, we are currently collaborating with political stakeholders on a real-world project to study the possible impacts of a low-carbon train within a rural transportation system of systems (Une étude sur l'écomobilité menée à Lectoure 2023).

The objective is to enable designers and decision-makers to represent scenarios of potential environmental and societal effects of design alternatives, in order to inform their design or strategy decisions. The ambition is to enable them to build their own model, to understand the sociotechnical dynamics, to get quick insights, and to be able to communicate it. The objective is to project scenarios of possible futures and to compare these scenarios relatively, with the help of indicators. Considering the uncertainty of the future, we propose a prospective approach: it is not a matter of predicting the most probable future, but of exploring several possible (and desirable) scenarios. To do so, we favor human understanding and intuition over quantitative data, building the model not only on quantitative data, but also on expert opinion, documentary studies, and mixed data collection methods (qualitative and quantitative surveys).

*Practical case of a second-hand clothing platform*

Let's take the simple example of a second-hand clothing platform. At first glance, we might imagine that it supports more sustainable practices and has a positive impact on society. However, it can have detrimental effects (Juge et al. 2022):

- Transporting clothes from seller to buyer and using the platform causes carbon emissions (first order).
- On average, buyers buy more items because they cost less or because they feel less guilty about buying second-hand (direct rebound).
- Some sellers use the money from sales to buy unnecessary new fast-fashion clothes (indirect rebound).
- Charities are suffering from a drop in donations because people are changing their habits and preferring to sell rather than donate (higher order rebound).

Identifying these effects can help define mitigation levers at the product design, service, and business model levels. For example, filtering the results to show only items that are close to the buyer, offering sellers the option of donating their item rather than selling it in certain cases, encouraging users to use the proceeds from sales on the platform through incentives, offering services to extend the lifespan of clothes through repair, and so on. However, this does not allow designers and decision-makers to determine which lever or combination of levers will have the most significant impact over time.

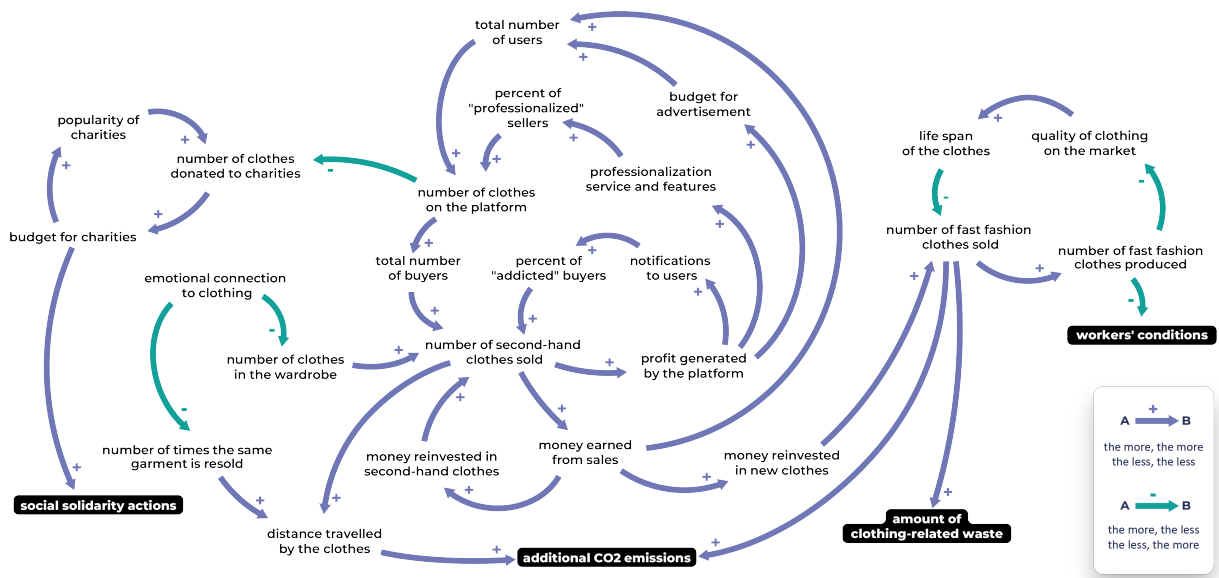

*Figure 1. Causal loop diagram (Kim, 1992) representing the influences between the different variables.*

As part of a project to redesign the platform, its business model and associated services, our methodology would consist of the following steps (this process is nonlinear and iterative):

1. Bring together stakeholders from the company and experts from the fashion and second-hand industries (including environmental economists and sociologists).
2. Collectively carry out a qualitative analysis of the various effects of the platform.
3. Collectively identify the variables of interest (e.g. CO2 emissions, amount donated to associations, etc.), and the influencing variables (e.g. number of garments sold, number of users, revenue generated by sales, etc.).
4. Collectively construct a diagram representing the influences between the different variables (see a simplified example in Figure 1).
5. Collectively define a strategy for quantifying these influences:
   a. by drawing on existing studies (e.g. average emissions per km travelled),
   b. by deducing from existing quantitative data measured on the platform (e.g. the average percentage of sales revenue used to buy other second-hand clothes on the platform),
   c. by carrying out a specific study (observations, qualitative interviews, quantitative surveys, etc.) with consumers (e.g. emotional link to second-hand clothes and number of clothes in the wardrobe),

d. by testing various hypotheses based on expert opinions.
6. Iteratively build a 'quali-quantitative' model using Magnitude, the prototype modeling tool (see the prototype in Figure 2), and explore several scenarios through simulation, not forgetting to represent possible rebound effects due to the mitigation measures.
7. Relatively compare the scenarios to collectively define a strategy at several levels (design, services, business), possibly including other stakeholders such as users, charities, etc.
8. Monitor the effects of this strategy over time, in comparison with the projected scenario, and iterate on the strategy and model based on observed changes over time.

This 'quali-quantitative' modeling is seen as a tool for collaborative thinking and decision support between different stakeholders. The relevance of the results depends on the model validity. However, the stakeholders are aware of the model's limitations since they have participated to its construction.

### *Magnitude: our prototype modeling tool*

To propose a simplified formalism that requires minimal or no coding, and to delve into the concept of a modeling tool tailored to interactions between product/service design and the associated sociotechnical system, we decided not to use generic system dynamics tools like InsightMaker, Stella, or Vensim. Instead, we developed our own prospective modeling tool (see Figure 2). For this tool, we opted to draw inspiration from the simplicity of causal loop diagrams and the calculation principles of stock-and-flow modeling.

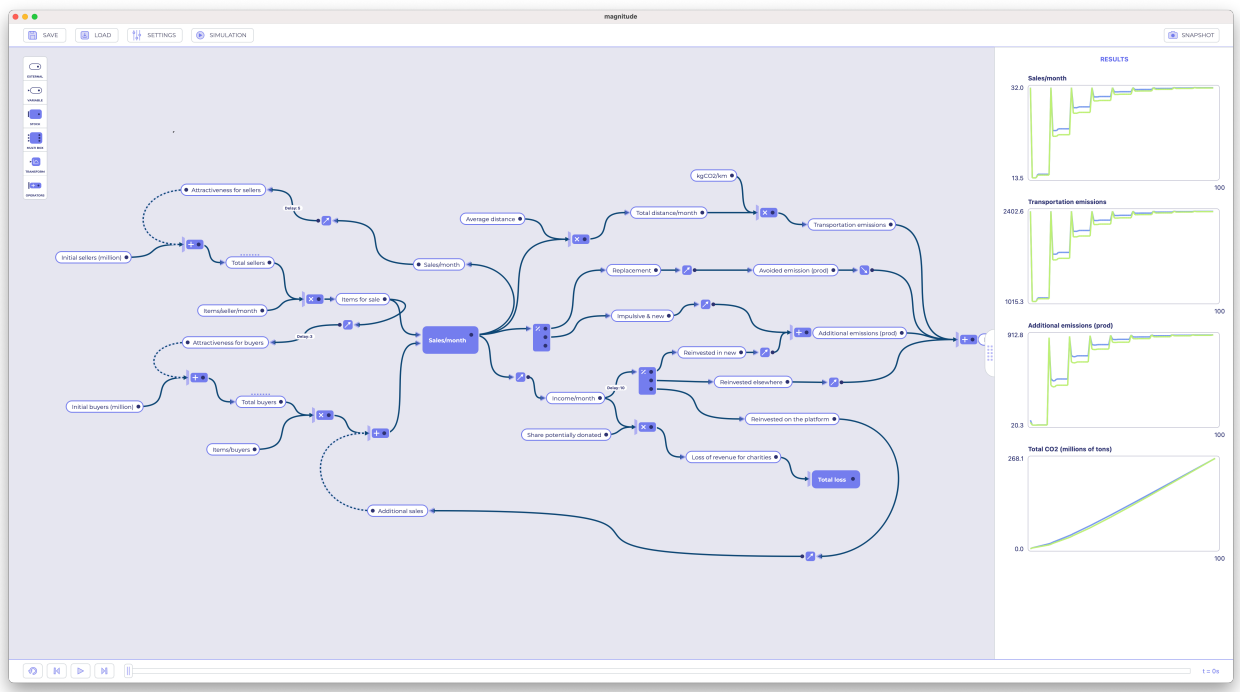

*Figure 2. Example of model building in the 'quali-quantitative' modeling tool*

Our prototype provides a toolbox that allows loading data from files, creating variables and stocks (cumulative variables) calculated at each time step, applying transformations and delays, and forming feedback loops. The curves of the values of variables of interest can be displayed on-demand in the results panel on the right. It is also possible to compare the curves of multiple simulations.

### *Perspectives*

The results of our initial practical cases seem promising, with good engagement from the involved designers. We also plan to apply this modeling methodology to other practical cases in the ICT (Information and Communication Technology) industry and compare the contribution in comparison to the methods and tools mentioned in Section 2.

*References*

Bérard, C. 2010. Group model building using system dynamics: An analysis of methodological frameworks. *Electronic Journal of Business Research Methods*, *8*(1), pp35-45.

Bornes, L. 2023. 'A Methodology and a Tool to Support the Sustainable Design of Interactive Systems: Adapting Systemic Design Tools to Model Complexity in Interaction Design'. In *Extended Abstracts of the 2023 CHI Conference on Human Factors in Computing Systems*, 1–5. Hamburg Germany: ACM. doi:10.1145/3544549.3577055.

Bornes, L., Letondal, C. and Vingerhoeds, R. 2022. 'Could Systemic Design Methods Support Sustainable Design Of Interactive Systems?'. In *Proceedings of Relating Systems Thinking and Design RSD11*. ISSN 2371-8404. https://rsdsymposium.org/could-systemic-design-methods-support-sustainable-design-of-interactive-systems

Bornes, L., Letondal, C. and Vingerhoeds, R. 2023. Using a Quali-Quantitative Modelling Tool to Explore Scenarios for More-Than-Sustainable Design. In *Proceedings of Relating Systems Thinking and Design RSD12*. https://rsdsymposium.org/quali-quantitative-modelling (pre-release)

Coroamă, V. C., P. Bergmark, M. Höjer, and J. Malmodin. 2020. 'A Methodology for Assessing the Environmental Effects Induced by ICT Services: Part I: Single Services'. In *Proceedings of the 7th International Conference on ICT for Sustainability*, 36–45. Bristol United Kingdom: ACM. doi:10.1145/3401335.3401716.

Coroamă, V.C. and Mattern, F. 2019. 'Digital rebound–why digitalization will not redeem us our environmental sins'. In *Proceedings 6th international conference on ICT for sustainability*. Lappeenranta. http://ceur-ws. org (Vol. 2382).

ITU (International Telecommunication Union). 2022. ITU-T L.1420. Enabling the Net Zero transition: Assessing how the use of information and communication technology solutions impact greenhouse gas emissions of other sectors.

Jevons, S. 1865. The Coal Question; An Inquiry Concerning the Progress of the Nation, and the Probable Exhaustion of Our Coal Mines. London, Macmillan and Co.

Jones, P. 2020. 'Systemic Design: Design for Complex, Social, and Sociotechnical Systems'. In *Handbook of Systems Sciences*, edited by Gary S. Metcalf, Kyoichi Kijima, and Hiroshi Deguchi, 1–25. Singapore: Springer Singapore. doi:10.1007/978-981-13-0370-8_60-1.

Juge, E., A. Pomiès, and I. Collin-Lachaud. 2022. 'Plateformes digitales et concurrence par la rapidité. Le cas des vêtements d'occasion'. Recherche et Applications en Marketing (French Edition) 37 (1): 37–60. doi:10.1177/0767370121994831.

Kim, D., H. 1992. 'Guidelines for Drawing Causal Loop Diagrams'. *The Systems Thinker* 3(1): 5–6.

Pathak, M., et al. 2021. 'Working group III contribution to the IPCC Sixth Assessment Report (AR6) – Technical Summary', *IPCC*, p. 132-133.

Raworth, K. 2017. Doughnut Economics: Seven Ways to Think Like a 21st Century Economist. Chelsea Green Publishing. ISBN:9781603586740.

Roussilhe, G. 2022. Les effets environnementaux indirects de la numérisation. Septembre 2022. https://gauthierroussilhe.com/articles/comprendre-et-estimer-les-effets-indirects-de-la-numerisation

Sterman, J. D. 2000. Business Dynamics—Systems Thinking and Modeling for a Complex World. Vol. 53. USA.

'Une Étude Sur l'écomobilité Menée à Lectoure'. 2023. *La Dépêche*. https://www.ladepeche.fr/2023/07/18/une-etude-sur-lecomobilite-menee-a-lectoure-11347331.php.

*About the authors*

Laetitia Bornes, PhD student, ENAC & ISAE-SUPAERO, www.linkedin.com/in/laetitia-bornes-design/, laetitia.bornes@isae-supaero.fr

After a master's degree in engineering, architecture and digital design, Laetitia Bornes worked for 5 years as a UX designer. During her PhD, she is exploring the potential of modeling to support the more-than-sustainable design of interactive systems. Her research aims to enable designers and decision-makers to represent the systemic effects of their design and explore scenarios to make informed decisions.

Catherine Letondal, Assistant Professor, ENAC, www.recherche.enac.fr/~letondal, catherine.letondal@enac.fr

Catherine Letondal has been conducting research into human-computer interaction and human-centred design for 25 years. She initially specialised in end-user development, participatory design and tangible interaction for complex critical domains such as aeronautics. For the past 5 years, she has been exploring HMI-based research directions for systemic approaches.

Rob Vingerhoeds, Professor, ISAE-SUPAERO, www.linkedin.com/in/rob-vingerhoeds-68523012/, rob.vingerhoeds@isae-supaero.fr

Rob Vingerhoeds is a Full Professor of Systems Engineering at ISAE-SUPAERO, Université de Toulouse, France. Rob's research interests include systems engineering and architecture, model-based systems engineering, concept design, and artificial intelligence techniques for engineering applications. Rob is Deputy Editor of the International Scientific Journal "Systems Engineering."